\documentclass[useAMS,usenatbib]{mn2e}

\usepackage[dvips]{graphicx}
\usepackage[cmex10]{amsmath}
\usepackage{array}
\usepackage{mdwmath}
\usepackage{mdwtab}

\title[A Bayesian technique for the detection of point sources in CMB maps]{A Bayesian technique for the detection of point sources in CMB maps}

\author[Arg\"ueso et al.]{F.~Arg\"ueso$^{1}$\thanks{E-mail:
    argueso@uniovi.es}, E. Salerno$^{2}$, D. Herranz$^{3}$, J. L.
    Sanz$^{3}$, E. E. Kuruo\u{g}lu$^{2}$ and K. Kayabol$^{2}$ \\\\
 $^{1}$ Departamento de Matem\'aticas, Universidad de Oviedo, 33007, Oviedo,
  Spain \\
$^{2}$  CNR Istituto di Scienza e Tecnologie dell'Informazione, via
G. Moruzzi 1, I-56124, Pisa, Italy \\ $^{3}$ Instituto de F\'\i sica
de Cantabria, CSIC-UC, Av. los Castros s/n, Santander, 39005, Spain}

\begin{document}

\date{Received --, Accepted --}

\pagerange{\pageref{firstpage}--\pageref{lastpage}} \pubyear{2010}

\maketitle

\label{firstpage}

\maketitle

\begin{abstract}
The detection and flux estimation of point sources in cosmic
microwave background (CMB) maps is a very important task in order to
clean the maps and also to obtain relevant astrophysical
information. In this paper we propose a maximum a posteriori (MAP)
approach detection method in a Bayesian scheme which incorporates
prior information about the source flux distribution, the locations
and the number of sources. We apply this method to CMB simulations
with the characteristics of the Planck satellite channels at 30, 44,
70 and 100 GHz. With a similar level of spurious sources, our method
yields more complete catalogues than the matched filter with a
$5\sigma$ threshold. Besides, the new technique allows us to fix the
number of detected sources in a non-arbitrary way.

\end{abstract}


\section{Introduction} \label{sec:intro}

The detection and estimation of the intensity of compact objects
embedded in a background plus instrumental noise is a problem of
interest in many different areas of science and engineering. A
classic example is the detection of point-like extragalactic objects
--i.e. galaxies-- in sub-millimetric Astronomy. Regarding this
particular field of interest, different techniques have proven
useful in the literature. Some of the existing techniques are: the
standard matched filter \citep[MF,][]{MF_radio92}, the matched
multifilter \citep{herr02a,lanz10} or the recently developed matched
matrix filters \citep{herranz08a}. Other methods include continuous
wavelets like the standard Mexican Hat \citep{wsphere} and other
members of its family \citep{MHW206}. All these filters have been
applied to real data of the Cosmic Microwave Background (CMB), like
those obtained by the WMAP satellite \citep{NEWPS07} and CMB
simulated data \citep{challenge08} for the experiment on board the
\emph{Planck} satellite \citep{planck_tauber05}. Besides, Bayesian
methods have also been recently developed \citep{hob03,psnakesI}. A
more detailed review on point source detection techniques in
microwave and sub-mm Astronomy, with a more complete list of
references, can be found in \cite{review2010}.

When a MF or a wavelet is applied to a CMB map in the blind
detection case, i.e. when it is assumed that the number of point
sources, their positions and fluxes are unknown, the most common
method for detection is based on the well-known idea of
thresholding: the maxima of the filtered map above a given threshold
are selected and considered as the positions of the sources, so that
the number of detected sources is the number of maxima above that
threshold. The fluxes are estimated then by using the corresponding
estimation formulas with the MF or the wavelet. The value of this
threshold remains arbitrary, though a $5\sigma$ cut is often
applied, since it guarantees that under reasonable conditions a few
detected sources are spurious. Apart from the arbitrariness of this
procedure, the prior knowledge regarding the average number of
sources in the surveyed patch, the flux distribution of these
sources or other properties are not used, so that useful information
is being neglected.

Bayesian detection techniques provide a natural way to take into account all the available information about the statistical distribution of both the sources and the noise.
Unfortunately, up to this date only a few works have addressed the problem of detecting extragalactic point sources in CMB data \citep{hob03,psnakesI}.
The reason for this is twofold: on the one hand, the statistical properties of extragalactic sources at sub-mm frequencies are still very poorly known.
On the other hand, mapping the full posterior probability density of the sources is often very difficult and computationally expensive.
These two problems explain, at least partially, the predominance of frequentist over Bayesian methods in the literature.
 Let us consider the previous two problems separately:

The microwave and sub-mm region has been until very recently one of
the last uncharted areas in astronomy. Concerning extragalactic
sources, this region of the electromagnetic spectrum is where the
total number of counts passes from being dominated by radio-loud
galaxies to being dominated by dusty galaxies. Although a minimum of
the emission coming from extragalactic sources is expected to occur
around 100--300 GHz, they are still considered as the main
contaminant of the CMB at small angular scales at these frequencies
\citep{tof98,zotti05}. The uncertainties about the number counts at
intermediate and low flux, redshift distribution, evolution and
clustering properties of this mixed population of objects are large.
In most cases this has motivated the use of noninformative priors,
which avoid to make adventurous assumptions about the sources but on
the other hand miss part of the power of the priors that are based
on observations and physical intuition.

But, in spite of what has been said above, our knowledge about the
statistical properties of point sources is growing day by day thanks
to the new generation of surveys and experiments. In the
high-frequency radio regime, WMAP observations are in agreement with
the de Zotti model \citep{zotti05,gnuevo08}. Priors for the number
density and flux distributions in the range of frequencies $ > 5 $
GHz are more and more reliable thanks to the information provided by
recent surveys such as CRATES at 8.4 GHz \citep{CRATES}, the
Ryle-Telescope 9C at 15.2 GHz \citep{taylor01,waldram03} or the
AT20G survey at 20 GHz \citep{ricci04,ATCA_BSS,ATCA10}. For a recent
review on radio and millimiter surveys and their astrophysical
implications, see \cite{review_dezotti10}. The situation is worse in
the far-infrared part of the spectrum, where relatively large
uncertainties remain in the statistical properties, the evolution
and, above all, the clustering properties of dusty galaxies. Most of
the existing dusty galaxy surveys have been carried out in the near
and medium infrared with IRAS, ISO and Spitzer, but the wave-band
from 60 to 500 $\mu$m is still virtually \emph{terra incognita}. The
only survey of a large area of the extragalactic sky at a wavelength
above 200 $\mu$m is the one recently carried out by the Herschel
pathfinder experiment, the Balloon Large Area Survey Telescope
\citep[BLAST,][]{BLAST}. In the next few months, however, the
luminosity function and the dust-mass function of dusty galaxies in
the nearby Universe will be much better understood thanks to the
Herschel-ATLAS Survey \citep{ATLAS}, which covers the wavelength
range between 110 and 500 $\mu$m and has already produced
interesting results during the Herschel Science Demonstration Phase
\citep{ATLAScounts}. Thanks to these and the previously mentioned
observations, the sub-mm gap is narrowing and our knowledge of
galaxy populations in this wave band, albeit far from perfect, is
quickly improving.

Apart from the uncertainties on the priors, the other complication that has traditionally deterred microwave astronomers from attempting Bayesian point source detection is computational and algorithmic complexity. Depending on the choice of priors and the likelihood function,
the full posterior distribution of the parameters of the sources may be very complex and in most cases it is impossible to obtain maximum a posteriori (MAP) values of the parameters and their associated errors via analytical equations. Numerical sampling techniques such as Monte Carlo Markov Chain (MCMC) methods are required in order to solve the inference problem, but these methods are computationally intensive. It is thus necessary to apply computing  techniques specifically tailored for accelerating the convergence and improving the efficiency of the sampling \citep{feroz08} and/or to find smart approximations of the posterior near its local maxima \citep{psnakesI}. But these enhancements have the cost of increasing dramatically the algorithmic complexity of the detection software, introducing new layers of intricacy in the form not only of additional assumptions and routines, but also of regularization 'constants', hidden variables,
hyperparameters and selection thresholds that in many cases must be fine-tuned manually in order to be adapted to the specific circumstances of a given data set. The complexity of the algorithms can rise to almost baroque levels, having a negative effect on the portability of the codes and on the reproducibility of the results.

We propose in this paper a simple strategy based on Bayesian
methodology which incorporates sensible prior information about the
source locations, the source fluxes and the source number
distribution. With these priors and assuming a Gaussian likelihood,
we can obtain an explicit form of the negative log-posterior of the
number of sources and their fluxes and positions. Assuming a MAP
methodology, we introduce a straightforward top-to-bottom detection
algorithm that allows us to determine the number, fluxes and
positions of the sources. We give a simple proof that the positions
of the sources \emph{must} be located in the local maxima of the
matched-filtered image if there is not a significant overlap between
sources. The main computational requirement of our algorithm is the
solution of a system of non-linear equations. Our method differs
from the one presented by \cite{psnakesI} in five main points:
\begin{enumerate}
  \item We use a more realistic set of priors for the source number, intensity and location distributions. In particular, our choice of the prior on the locations is also flat but depends on the number of sources $n$, which later proves to be decisive for the log-posterior.
  \item We obtain an explicit form of the negative log-posterior and an explicit solution of the MAP estimate of the source intensities as the solution of a non-linear system of equations.
  \item We prove that, for non-overlapping sources and a Gaussian likelihood, the MAP estimation of the positions of the sources is given by the location of the local maxima of the matched filtered images.
  \item Since we are interested only in point sources, we fix the size parameter of the objects to be detected.
  \item We can also find the MAP solution for the number of sources present in the images with a simple top-to-bottom search strategy. We do not need to resort to costly evaluations of the Bayesian evidence.
\end{enumerate}

The layout of the paper is as follows: in
section~\ref{sec:methodology} we present the method and derive the
corresponding posterior which includes the data likelihood and the
priors. In section~\ref{sec:simulations} we apply the method to CMB
simulations with the characteristics of the radio Planck channels
(from 30 to 100 GHz, where the number count priors are most
reliable) and compare it with the standard procedure of using a MF
with a $5\sigma$ threshold. The main results are also presented in
section~\ref{sec:simulations}. The conclusions are given in the
final section.

\section{Methodology} \label{sec:methodology}

In a region of the celestial sphere, we suppose to have an unknown
number $n$ of radio sources that can be considered as point-like
objects if compared to the angular resolution of our instruments.
This means that their actual size is smaller than our smallest
resolution cell. The emission of these sources is superimposed to a
radiation $f(x,y)$ coming from diffuse or extended sources. In our
particular case this radiation is the CMB plus foreground radiation.
A model for the emission as a function of the position $(x,y)$ is
\begin{equation} \label{eq:model}
\tilde{d}(x,y)=f(x,y)+\sum_{\alpha=1}^{n}
a_{\alpha} \delta(x-x_{\alpha},y-y_{\alpha}),
\end{equation}
\noindent where $\delta (x, y)$ is the 2D Dirac delta function, the
pairs $ (x_{\alpha},y_{\alpha}) $ are the locations of the point
sources in our region of the celestial sphere, and $a_{\alpha}$  are
their fluxes. We observe this radiation through an instrument, with
beam pattern $b(x,y)$, and a sensor that adds a random noise
$n(x,y)$ to the signal measured. Again, as a function of the
position, the output of our instrument is:
\begin{equation} \label{eq:output}
d(x,y)=\sum_{\alpha=1}^{n}a_{\alpha} b(x-x_{\alpha},y-y_{\alpha})+(f*b)(x,y)+n(x,y),
\end{equation}
\noindent where the point sources and the diffuse radiation have
been convolved with the beam. In our application, we are interested
in extracting the locations and the fluxes of the point sources. We
thus assume that the fluxes of the point sources are sufficiently
above the level of the rest of the signal plus the noise, and
consider the latter as just a disturbance superimposed to the useful
signal. If $\epsilon (x,y)$ is the sum of the diffuse signal plus
the noise, model (\ref{eq:output}) becomes
\begin{equation} \label{eq:output2}
d(x,y)=\sum_{\alpha=1}^{n}a_{\alpha} b(x-x_{\alpha},y-y_{\alpha})+\epsilon
(x,y).
\end{equation}
\noindent If our data set is a discrete map of $N$ pixels, the above
equation can easily be rewritten in vector form, by letting
$\textbf{d}$ be the lexicographically ordered version of the
discrete map $d(x,y)$, $\textbf{a}$ be the $n$-vector containing the
positive source fluxes $a_{\alpha}$, $\boldsymbol{\epsilon}$ the
lexicographically ordered version of the discrete map,
$\epsilon(x,y)$, and $\boldsymbol{\phi}$ be an $N\times n$ matrix
whose columns are the lexicographically ordered versions of $n$
replicas of the map $b(x,y)$, each shifted on one of the source
locations. Equation (\ref{eq:output2}) thus becomes
\begin{equation} \label{eq:output_vec}
\textbf{d}=\boldsymbol{\phi}  \textbf{a}+ \boldsymbol{\epsilon}.
\end{equation}

Looking at equations (\ref{eq:output2}) and (\ref{eq:output_vec}),
we see that, if the goal is to find locations and fluxes of the
point sources, our unknowns are the number $n$, the list of
locations $(x_{\alpha},y_{\alpha})$, with $\alpha=1,\ldots, n$ and
the vector $\textbf{a}$. It is apparent that, once $n$ and
$(x_{\alpha},y_{\alpha})$  are known, matrix $\boldsymbol{\phi}$ is
perfectly determined. Let us then denote the list of source
locations by the $n\times 2$ matrix $\textbf{R}$, containing all
their coordinates. If we want to adopt a Bayesian strategy to solve
our problem, we must be able to write the posterior probability
density of our unknowns. A suitable estimation criterion must then
be chosen.

\subsection{Posterior} \label{sec:posterior}

By the Bayes rule, the posterior we are looking for has the
following form
\begin{equation} \label{eq:posterior}
p(n,\textbf{R},\textbf{a}|\textbf{d})\propto
p(\textbf{d}|n,\textbf{R},\textbf{a}) p(n,\textbf{R},\textbf{a})
\end{equation}
\noindent where $p(\textbf{d}|n,\textbf{R},\textbf{a})$  is the
likelihood function, derived from our data model
(\ref{eq:output_vec}). To find the prior density
$p(n,\textbf{R},\textbf{a})$ we need to make a number of
assumptions. Let us first observe that, in principle, both
$\textbf{R}$ and $\textbf{a}$ depend on $n$, through the number of
their elements. On the other hand, we can safely assume that, once
$n$ is fixed, the fluxes $\textbf{a}$ of the sources are independent
of their locations. These assumptions lead us to write
\begin{equation} \label{eq:separated_posterior}
p(n,\textbf{R},\textbf{a})=p(\textbf{R},\textbf{a}|n)p(n)=p(\textbf{R}|n)p(\textbf{a}|n)p(n)
\end{equation}

\noindent This expression is valid when we consider extragalactic
point sources, whose fluxes are not related to their positions. This
will be the case in this paper.

\subsection{Likelihood function}

As mentioned above, the likelihood function derives from the physics
associated to the assumed data model. In general, unfortunately, a
data model of type (\ref{eq:output}) is difficult to describe
statistically. We are going to assume from now on that
$\boldsymbol{\epsilon}$ is a random Gaussian field with zero mean
and known covariance matrix $\xi$. This is true if we only consider
the CMB and the instrumental noise, excluding other foregrounds. In
this paper we will deal with zones of the sky where the foreground
contribution is not important or where the foregrounds have been
conveniently removed by component separation techniques. The
likelihood is thus
\begin{equation} \label{eq:likelihood}
p(\textbf{d}|n,\textbf{R},\textbf{a})\propto \exp
\left[-\frac{(\textbf{d}-\phi \textbf{a})^t \boldsymbol{\xi}^{-1}
(\textbf{d}-\phi \textbf{a})}{2} \right].
\end{equation}
\noindent Observe that the negative of the exponent in
(\ref{eq:likelihood}) is in any case the squared
$\boldsymbol{\xi}^{-1}$-norm fit of the reconstructed data to the
measurements, and this always carries information about the goodness
of our estimate. However, if the Gaussian assumption is not
verified, function (\ref{eq:likelihood}) is not the likelihood of
our parameters, and when it is introduced in (\ref{eq:posterior}),
we do not obtain the posterior distribution we are looking for. In
our simulations we will also include the confusion noise due to
faint extragalactic sources. This confusion noise is not Gaussian
but as we will see later, it does not hamper the detections, since
its standard deviation is much lower than that of the CMB plus the
instrumental noise for the frequencies considered in this paper. For
the sake of simplicity, we will defer to further papers the
treatment of the more general case which includes the foregrounds.

\subsection{Prior on source locations} \label{sec:prior_locations}

A priori, it is reasonable to assume that all the different
combinations of $n$ distinct locations occur with the same
probability. Then function $p(\textbf{R}|n)$ in Eq. (\ref{eq:separated_posterior}) can be
written as
\begin{equation} \label{eq:prior_locations}
p(\textbf{R}|n)=\displaystyle\frac{n!(N-n)!}{N!},
\end{equation}
\noindent since $N!/(n!(N-n)!)$ is the number of possible distinct
lists of $n$ locations in a discrete $N$-pixel map. This assumption
is based on the fact that the sources considered in this paper are
spatially uncorrelated.

\subsection{Prior on fluxes} \label{sec:prior_amplitudes}

Experimentally, it has been found that the fluxes of the strongest
sources are roughly distributed as a negative power law, with
exponent $\gamma$. Conversely, the weak sources have fluxes that are
roughly uniformly distributed. To include these two behaviors into a
single formula, one should first discriminate in some way between
weak and strong sources. This can be done empirically, by
establishing a sort of threshold $a_0$ on the fluxes and a
conditional prior with the form of the Generalized Cauchy
Distribution \citep{rider57}:
\begin{equation} \label{eq:gcauchy}
p(\textbf{a}|n)\propto \prod _{\alpha=1}^{n}
\left[1+\left(\frac{a_{\alpha}}{a_0}\right)^p\right]^{-\frac{\gamma}{p}},
\end{equation}
\noindent with $p$ a positive number. Distribution
(\ref{eq:gcauchy}) obviously assumes that the fluxes of the
different sources are mutually independent. This prior clearly shows
the behavior required for strong and weak sources. In order to work
with non-dimensional quantities, we define
$x_{\alpha}=a_{\alpha}/a_0$; we also assume that we will detect
point sources above a minimum flux $a_m$, that leads to the
following normalized distribution
\begin{equation} \label{eq:normalized_gcauchy}
p(\textbf{x}|n)=\prod _{\alpha=1}^{n}
\frac{p}{B\left(\frac{1}{1+x_m^p};\frac{\gamma-1}{p},\frac{1}{p}\right)}
\left(1+x_{\alpha}^p\right)^{-\frac{\gamma}{p}},\quad
x_{\alpha}\epsilon [x_m,\infty)
\end{equation}
\noindent where $B$ is the incomplete beta function and
$x_m=a_m/a_0$. In the next section, we determine the values of
$a_0$,  $p$ and $\gamma$, by fitting this formula to the point
source distribution given by the de Zotti counts model
\citep{zotti05}.

\subsection{Prior on the number of sources} \label{sec:prior_number}

We need to establish a discrete probability distribution that
expresses the probability of a number of occurrences in a fixed
domain once their average density is known, their locations in the
domain are mutually independent, and no pair of sources can occur in
the same location. All these assumptions seem reasonable when
applied to the configurations of the point sources in the celestial
sphere, at least for radio-frequencies \citep{argueso03,gnuevo05}.
Assuming a continuous map domain, all the requirements mentioned are
satisfied by the Poisson distribution. Strictly speaking, we have a discrete
$N$-pixel map, so a binomial distribution should be used, but if $N$ is
not too small, a Poisson distribution should model correctly the
probability of having $n$ occurrences of point sources. The prior on $n$
appearing in (\ref{eq:separated_posterior}) is thus
\begin{equation} \label{eq:prior_number}
p(n)=\frac{\lambda^n e^{-\lambda}}{n!},
\end{equation}
\noindent where $\lambda$, the intensity of the Poisson variable, is
the expected number of sources in the map at hand. The value of
$\lambda$ will depend on the flux detection limit $a_m$, the size of
the map and the wavelength of the observation.

\subsection{An explicit expression for the negative log-posterior} \label{sec:log_posterior}

If we multiply all the factors which appear in (\ref{eq:posterior}) and (\ref{eq:separated_posterior}) and
calculate the negative log-posterior, we find (apart from additive
constants)
\begin{eqnarray} \label{eq:log_posterior}
L(n,\textbf{R},\textbf{x}) & = & \frac{1}{2} \left(\textbf{x}^t
\mathbf{M} \textbf{x}-2\textbf{e}^t \textbf{x} \right)- \log(N-n)!
- n \log(\lambda) \nonumber \\
 & - & n \log(p)+ n \log
B\left(\frac{1}{1+x_m^p};\frac{\gamma-1}{p},\frac{1}{p}\right) \nonumber \\
 & + & \frac{\gamma}{p}\sum_{\alpha=1}^{n}
\log \left(1+x_{\alpha}^p \right),
\end{eqnarray}
\noindent with $\mathbf{M}=a_0^2 \boldsymbol{\phi}^{t}
\boldsymbol{\xi}^{-1} \boldsymbol{\phi}$ and  $\mathbf{e}=a_0
\boldsymbol{\phi}^{t}\boldsymbol{\xi}^{-1}\mathbf{d}$. The
correlation matrix $\boldsymbol{\xi}$ is computed by using the
$C_{\ell}$'s obtained from the WMAP five-year maps \citep{Nolta09}
and adding the instrumental noise. We assume that we know $a_0$,
$p$, $x_i$, $\gamma$ and $\lambda$, in fact we calculate them by
using the de Zotti counts model \citep{zotti05}. Therefore, the
unknowns are: the normalized fluxes $\textbf{x}$, the number of
point sources $n$ and the positions of the point sources through the
matrix $\boldsymbol{\phi}(\mathbf{R})$.

Let us now examine the structure of function
(\ref{eq:log_posterior}). The first term comes from the likelihood,
and obviously decreases as much as our solution fits the data. The
second term takes into account the prior for the source
configuration and penalizes large values of n. The third term comes
from the prior on the number of sources and, depending on the value
of $\lambda$, favors ($\lambda >1$) or disfavors ($\lambda<1$) the
increase of the number of sources. The  following terms come from
the prior on the source fluxes conditioned to $n$, the last one
introduces an additional cost as soon as a new source is added to
the solution. If $N \gg \lambda$ and $N \gg n$, what is typical in
CMB maps, it can be proven by using Stirling's approximation that
the second term is the dominant one coming from the priors. In the
next section, we will analyze with simulations the contribution of
each particular term.

\subsection{Maximum a posteriori (MAP) solution} \label{sec:maxposterior}

Formula (\ref{eq:log_posterior}) includes all the information about
the positions, fluxes and number of sources. In order to obtain
concrete results, we will choose the values of $\textbf{R}$,
$\textbf{x}$ and $n$ which maximize the posterior. This choice will
be justified by means of the simulations and results that we will
present in the next section.

Therefore, regarding the flux we minimize (\ref{eq:log_posterior})
with respect to $\textbf{x}$, by taking the derivative and equating
to zero and we obtain
\begin{equation} \label{eq:solution1}
\sum_{\beta=1}^{n} M_{\alpha \beta}
x_{\beta}-e_{\alpha}+\frac{\gamma
x_{\alpha}^{p-1}}{1+x_{\alpha}^{p}}=0.
\end{equation}
\noindent By solving (\ref{eq:solution1}) numerically we would
obtain the estimator of $\textbf{x}$ which yields the maximum
posterior probability. However, we know neither the number of
sources nor their positions. In order to determine the positions, we
assume that the point sources are in the local maxima of
$\textbf{e}$, which is the matched-filtered map of the original
data. In the following, we will show that this assumption can be
safely adopted, since the minima of $L$ must be in the maxima of the
matched-filtered map (we also remark that the matched filter is not
introduced ad hoc, but it appears naturally as a part of the
formalism).

For simplicity, let us assume that we have only one source, in this
case the terms of $L$ which depend on the flux can be written
\begin{equation} \label{eq:solution_1s}
L(x)=\frac{M_{11}x^2}{2}-e_1x+\frac{\gamma}{p} \log\left(1+x^p\right),
\end{equation}
\noindent where $M_{11}$ and $e_1$ are the corresponding values of
$\mathbf{M}$ and $\textbf{e}$ at the pixel supposedly occupied by
the point source. If we take the derivative of
(\ref{eq:solution_1s}) with respect to $x$ and equate to zero, we
obtain the following equation for $\hat{x}$, the estimator of $x$:
\begin{equation} \label{eq:estimatex_1s}
M_{11}\,x+\frac{\gamma x^{p-1}}{1+x^p}=e_1 \Rightarrow
\hat{x}=\hat{x}(e_1).
\end{equation}
\noindent If we substitute the last expression in eq.
(\ref{eq:solution_1s}), we can write the expression for the negative
log-posterior $L(\hat{x}(e_1))$
\begin{equation}\label{eq:detailedlike}
L(\hat{x}(e_1))=-\frac{M_{11}\hat{x}^2}{2}-\frac{\gamma
\hat{x}^{p}}{1+\hat{x}^p}+\frac{\gamma}{p}
\log\left(1+\hat{x}^p\right).
\end{equation}

 \noindent By taking the derivative of this formula with respect to $e_1$, we
finally find
\begin{equation} \label{eq:justification}
\frac{dL}{de_1}= \frac{dL}{d\hat{x}}
\left(\frac{de_1}{d\hat{x}}\right)^{-1}=-\hat{x}(e_1).
\end{equation}
\noindent where we have calculated the derivative in
(\ref{eq:estimatex_1s}). Since the estimated value of the flux must
be positive, this expression shows that the negative log-posterior
at the estimated value of $x$ decreases with $e_1$, so it is minimum
at the highest value of $e_1$ i.e. at the maximum of the
matched-filtered map. Therefore, the posterior, calculated at the
estimated flux value, is maximum when we assume that the point
source is at a peak of the map. This conclusion is valid if we have
more than one source, provided that there is no overlap between
sources, i.e. the areas where the individual images of all the
sources are nonzero must be completely disjoint, because in this
case each source can be treated individually.

In order to determine the number of sources, we sort these local
peaks from top to bottom and solve (\ref{eq:solution1}) successively
adding a new source. At the same time, we calculate the negative
log-posterior (\ref{eq:log_posterior}) and choose the number $n$ of
sources which produce the minimum value of (\ref{eq:log_posterior}).
In this way we have constructed an objective stopping criterion
which yields, by combining (\ref{eq:log_posterior}) and
(\ref{eq:solution1}), the number of sources and their fluxes which
maximize the posterior. In the next section, we apply the method
explained above to the detection and flux estimation of point
sources in CMB maps.

In order to compare this technique with a standard method, we also
calculate the local peaks above a certain threshold, for instance a
$5 \sigma$ threshold, and solve (\ref{eq:solution1}) with
$\gamma=0$, that amounts to using a MF, i.e. a maximum likelihood
estimator.

\section{Simulations and results} \label{sec:simulations}

\subsection{The simulations} \label{sec:sim_description}

In order to check the performance of the new technique, we have
carried out simulations including CMB, instrumental noise and point
sources. The simulations have the characteristics of the 30, 44, 70
and 100 GHz channels of the Planck satellite: pixel size, beam width
and instrumental noise\footnote{For details on the Planck
instrumental and scientific performance, see the Planck web site
 http://www.rssd.esa.int/index.php?project=PLANCK.}. The simulations are
flat patches of $32\times 32$ pixels (30 and 44 GHz) and $64\times
64$ pixels (70 and 100 GHz), so that the size of each patch is
$3.66\times 3.66$ square degrees. In order to avoid border effects,
we simulate patches of four times this size and keep the central
part for our analysis. The small size of the simulations allows us
to do our calculations in a fast way. We perform 1000 simulations
for each channel.

The CMB maps have been generated by using the power spectrum, the
$C_{\ell}$'s, that produces the best fit to the WMAP five-year maps
\citep{Nolta09}, we have also added the instrumental noise of the
30, 44, 70 and 100 GHz channels of the Planck satellite. Finally, we
have simulated point sources, by taking into account the flux
distribution predicted by the de Zotti model \citep{zotti05}. We
have included the faint tail of the de Zotti distribution,
simulating point sources from $0.01 $ mJy on. In this way, we have
considered the confusion noise due to unresolved point sources. The
standard deviation of this confusion noise is much lower than that
of the CMB plus instrumental noise in these channels.

\begin{figure}
\includegraphics[width=\columnwidth]{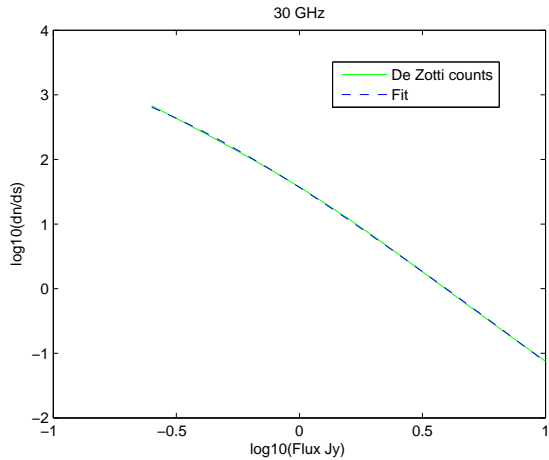}
 \caption{\label{fig1}{
 $\log_{10}$ of the differential counts plotted against the flux for the de Zotti model (green line) and the fit to the extended power-law given by (\ref{eq:gcauchy}) with the parameters of Table~\ref{tb:table1} (blue line).
 The two lines are nearly indistinguishable.  }}
\end{figure}

For each simulation we consider the negative log-posterior given by
(\ref{eq:log_posterior}). In this equation we see several
magnitudes, $\gamma$, $a_0$, $a_m$ and $\lambda $, which depend on
the frequency. By fitting the number counts given in the de Zotti
model by (\ref{eq:gcauchy}), we calculate $\gamma$ and $a_0$. We
have taken $p$=1 in (9), since the goodness-of-fit obtained by
changing $p$ is not better than that of the particular case $p=1$.
The parameter $\lambda$ is the average number of sources per patch
and $a_m$ is the minimum flux that we consider in our detection
scheme, we have chosen $a_m=0.25$ Jy, the typical rms deviation of
the CMB plus noise maps at the frequencies we consider. The values
of $\gamma$, $a_0$ and $\lambda $ are shown in Table~\ref{tb:table1}
for the different frequencies. In Figure~\ref{fig1} we show as an
example the fit to our extended power-law (\ref{eq:gcauchy}) in the
case of the 30 GHz channel. It is clear that the extended power-law
fits very well the counts predicted by the de Zotti model. The value
of $\chi^2$ is $(2-3)\times 10^{-3}$ giving probabilities very close
to $1$.

\begin{table}
  \centering
  \begin{tabular}{cccc}
  \textbf{frequency} & \textbf{$ \gamma $} & \textbf{$ a_0 $} & \textbf{ $\lambda $} \\
  \hline\hline
  \textbf{30 GHz} &2.90 & 0.19& 0.69  \\
    \textbf{44 GHz} & 2.87  & 0.15  & 0.53  \\
    \textbf{70 GHz} &2.87  & 0.15  & 0.49   \\
    \textbf{100 GHz} & 2.87& 0.15  & 0.47  \\
    \hline\hline
\end{tabular}
  \caption{ Values of the power-law exponent $\gamma$ and the flux $ a_0$ in Jy, as obtained by fitting the De Zotti counts.
  $ \lambda$ is the average number of point sources above $0.25$ Jy
  in the considered patches.
  }\label{tb:table1}
\end{table}

\subsection{Discussion on the performance of the algorithm}

For each simulation we calculate
$\mathbf{M}=a_{0}^{2}\boldsymbol{\phi}^t \boldsymbol{\xi}^{-1}
\boldsymbol{\phi}$ and $\textbf{e}=a_0 \boldsymbol{\phi}^t
\boldsymbol{\xi}^{-1} \textbf{d}$. We obtain $\boldsymbol{\xi}^{-1}$
from the WMAP $C_{\ell}$'s, taking into account the effects of the
pixel and the beam windows and the corresponding noise levels for
each channel.  We calculate the estimated fluxes $\hat{x}_{\alpha}$
by solving (\ref{eq:solution1}) as explained in the previous
section: we select the maxima of $\textbf{e}$ above a certain
threshold (we choose a $1\sigma$ threshold so that we have a
suitable number of peaks) and we perform a top to bottom strategy,
i.e. we sort the local maxima downwards from higher to lower values
and starting from the highest peak we solve (\ref{eq:solution1})
including in each new iteration a new local maximum. At the same
time, we calculate (\ref{eq:log_posterior}) and stop the iterations
when we find the minimum value of the negative log-posterior. In
this way, we obtain the source fluxes and the number of sources that
maximize the log-posterior. We also calculate the local peaks of
$\textbf{e}$ above a $5\sigma$ threshold, a standard detection
method, and calculate the source flux by solving
(\ref{eq:solution1}) with $\gamma=0$, this is equivalent to using a
MF with a $5\sigma$ threshold. Our intention is to compare the
Bayesian method (BM), with prior information and a natural stopping
criterion, and the standard MF.

\begin{figure}
\includegraphics[width=\columnwidth]{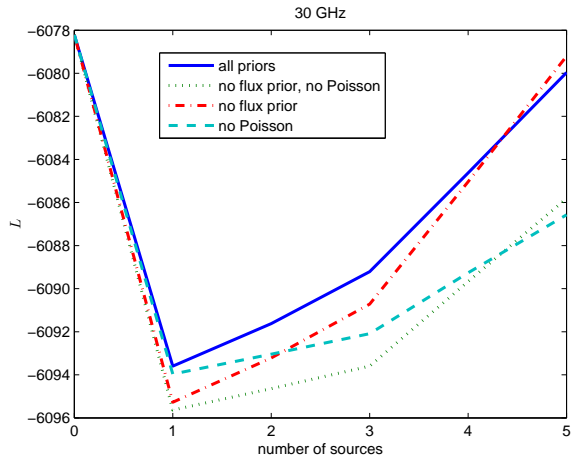}  
 \caption{\label{fig2}{
  Negative log-posterior against the number of detected sources for a simulation at 30 GHz. We have included in the posterior: all the priors (blue solid line), all the priors but the source flux distribution
 (red dash-dotted line), all the priors but the Poisson source number distribution (cyan dashed line) and finally we have excluded both the Poisson and the source flux distribution (green dotted line).    }}
\end{figure}

According to our simulations, the fundamental contributions to the
posterior come from the likelihood (\ref{eq:likelihood}) and the
prior on source locations (\ref{eq:prior_locations}). The other
terms also contribute, but as can be seen in Figure~\ref{fig2},
where we show the negative log-posterior plotted against the number
of sources for a particular simulation, their influence is not so
important. The likelihood tends to increase the number of detected
sources, over-fitting the data and the prior
(\ref{eq:prior_locations}) tends to decrease the number of sources.
The combination of (\ref{eq:likelihood}) and
(\ref{eq:prior_locations}) fixes the most probable number of point
sources, though the other two terms, although less important can
have some influence. This shows the robustness of the method with
respect to small changes in the parameters of priors (10) and (11).

\begin{figure}
\includegraphics[width=\columnwidth]{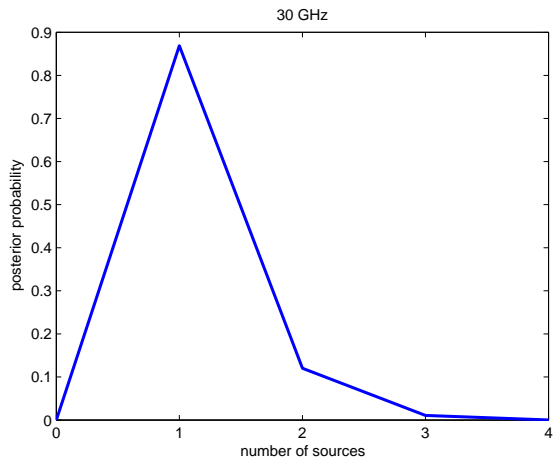}
 \caption{\label{fig3}{ Posterior probability against number of detected point sources for a simulation at the 30 GHz channel with one real source.    }}
\end{figure}

We also raise the question whether the estimated number of point
sources gives us a clearly higher posterior probability, i.e
$\propto \exp(-\log L) $ than other close numbers. The answer can be
seen in Figure~\ref{fig3}, where we show, as an example, the
normalized posterior probability plotted against the number $n$ of
point sources for a given simulation with one real source. The
probability is clearly peaked at the estimated number of sources,
which is the real number of point sources in this case. In our
simulations we observe that the posterior probability is always
strongly peaked around the estimated number of sources.

We analyze 1000 simulations for each of the considered Planck
channels: we calculate the contamination (the number  of detected
spurious sources over the total number of detected sources above a
given flux), the completeness (the number of real detected sources
over the number of simulated sources above a given flux) and the
average of the absolute value of the relative error of the estimated
flux with respect to the real flux (reconstruction error). We count
a detected source as real when there is a real simulated source at a
distance no longer than two pixels from the detected one, this
distance is the position error. This real source must have a flux
equal or higher than 0.20 Jy, to fix a threshold close to the
$1\sigma $ level of the CMB plus noise map. The same conditions are
required for the MF.

In order to give the uncertainty in the flux, derived from our
Bayesian approach, we will obtain a $95\%$ confidence interval
associated to our probability distribution
\begin{equation} \label{eq:probability}
P(x)\propto \exp(-L(x)),
\end{equation}

\noindent where $L(x)$ is given by eq. \ref{eq:solution_1s}. This
will be called the estimation error. We can also calculate the
expectation value of the flux from this distribution, this value can
be compared with the most probable value (i.e. our estimated flux).

\subsection{Results}

Taking into account the considerations above, we have applied our algorithm to the simulations described in~\ref{sec:sim_description} and obtained the following results.

\begin{figure}
\includegraphics[width=\columnwidth]{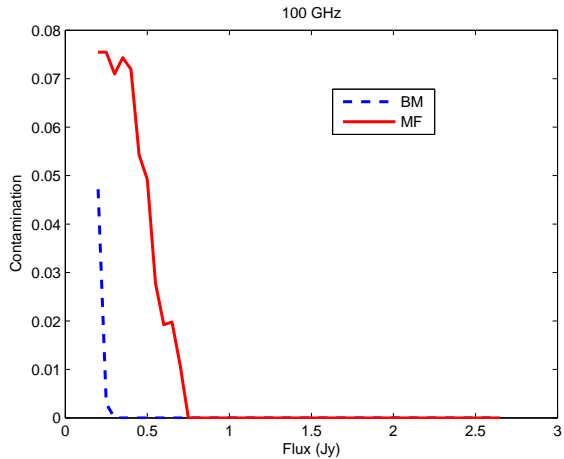}
 \caption{\label{fig4}{
 Contamination plotted against the flux for the BM and the MF (100 GHz).    }}
\end{figure}

At the 30 GHz channel we have $4.9\%$ contamination
 above 0.2 Jy with the BM, and $0.7\%$
with the MF. However, the completeness is much better for the BM,
$64\%$, than for the MF, $22\%$. From 0.7 Jy on we do not have any
spurious source (BM) and the completeness is $99\%$. For the MF
there are no spurious sources from 0.25 Jy on, but the completeness
at 0.7 Jy is only $75\%$. In regard to the average value of the
absolute value of the relative error (reconstruction error), when we
calculate this error in flux intervals of 0.1 Jy, we obtain similar
values for the BM and the MF. For instance, we obtain errors below
$15\%$ from $0.6$ Jy on and below $10\%$ from 1 Jy on for both
methods. Only $14\%$ of the sources have a reconstruction error on
the position of 1 pixel and only $3\%$ have a higher error. These
results are nearly the same for the BM and the MF.

\begin{figure}
\includegraphics[width=\columnwidth]{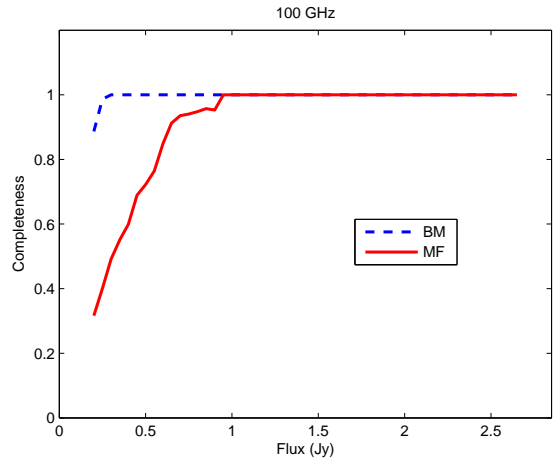}
 \caption{\label{fig5}{
 Completeness plotted against the flux for the BM and the MF (100 GHz).    }}
\end{figure}

At the 44 GHz channel we have $6.5\%$ contamination for fluxes
higher than 0.2 Jy with the BM, and  $4\%$ with the MF. The
completeness is
 $37\%$ for the BM and $11\%$ for the MF. From 0.8 Jy on we do not have any
spurious source (BM) and the completeness is $100\%$. For the MF
there are no spurious sources from 0.30 Jy on, but the completeness
at 0.8 Jy is only $70\%$. As in the 30 GHz case, we obtain similar
average values of the absolute value of the relative error for both
methods. For instance, we obtain errors below $15\%$ from 0.90 Jy on
in both cases. $15\%$ of the sources have a reconstruction error on
the position of 1 pixel and only $3\%$ have a higher error. These
results are nearly the same for the BM and the MF.

\begin{figure}
\includegraphics[width=\columnwidth]{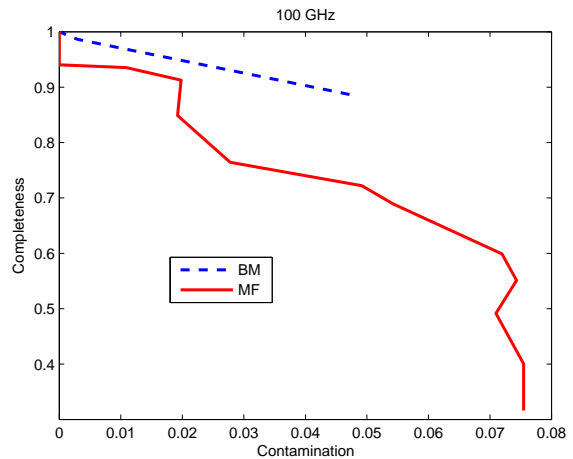}
 \caption{\label{fig6}{
 Completeness plotted against contamination for the BM and the MF (100 GHz).    }}
\end{figure}

At the 70 GHz channel we have  $3.2\%$ of spurious sources for
fluxes higher than 0.2 Jy with the BM, and $1\%$ with the MF. The
completeness is $45\%$ for the BM and $19\%$ for the MF. From 0.45
Jy on we do not have any spurious source (BM) and the completeness
is $96\%$. For the MF there are no spurious sources from 0.30 Jy on,
but the completeness at 0.45 Jy is only $46\%$. As in the cases
above, we obtain similar average values of the absolute value of the
relative error for both methods. For instance, we obtain errors
below $15\%$ from 0.60 Jy on in both cases. $9\%$ of the sources
have a reconstruction error on the position of 1 pixel and only
$1\%$ have a higher error. These results are similar for the BM and
the MF.

At the 100 GHz channel we have $4.7\%$ of spurious sources for
fluxes higher than 0.2 Jy with the BM, and $7.5\%$ with the MF. The
completeness is $89\%$ for the BM and $32\%$ for the MF. From 0.3 Jy
on we do not have any spurious source (BM) and the completeness is
$100\%$. For the MF there are no spurious sources from 0.75 Jy on
and the completeness at 0.75 Jy is $94\%$. As in the cases above, we
obtain similar average values of the absolute value of the relative
error for both methods. For instance, we obtain errors below $10\%$
from 0.5 Jy in both cases. $1\%$ of the sources have a
reconstruction error on the position of 1 pixel and there are no
sources with a higher error. These results are similar for the BM
and the MF.

\begin{figure}
\includegraphics[width=\columnwidth]{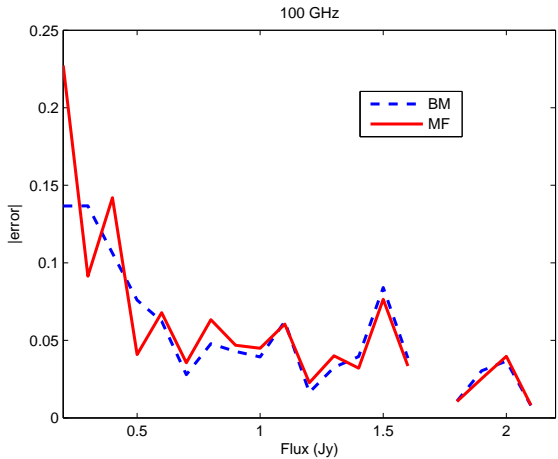}
 \caption{\label{fig7}{
 Average value of the absolute value of the relative error plotted against the flux for the BM and the MF (100 GHz). We can see in the plot the low values of the error for both methods.}}

\end{figure}

\begin{figure}
\includegraphics[width=\columnwidth]{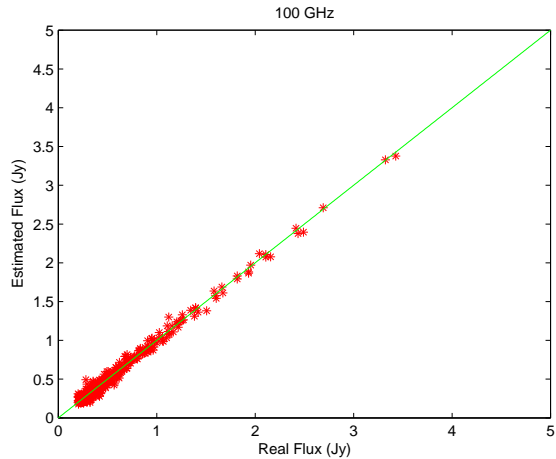}
 \caption{\label{fig8}{
 Estimated flux against real flux for the BM (100 GHz). We have plotted the straight line $y=x$ for comparison.    }}
\end{figure}

In order to visualize these results, we have plotted for the 100 GHz
channel the contamination (integrated contamination) against the
flux in Figure~\ref{fig4}, the completeness (integrated
completeness) against the flux in Figure~\ref{fig5}, the
completeness against the contamination in Figure~\ref{fig6}, the
average value of the absolute value of the relative error in
Figure~\ref{fig7} and finally, the estimated flux against the real
flux in Figure~\ref{fig8}. It is clear that for a given value of the
contamination the completeness is higher for the BM than for the MF.
Although the results are similar at all the studied frequencies, we
have chosen the 100 GHz channel in order not to complicate
unnecessarily the figures.

\begin{figure}
\includegraphics[width=\columnwidth]{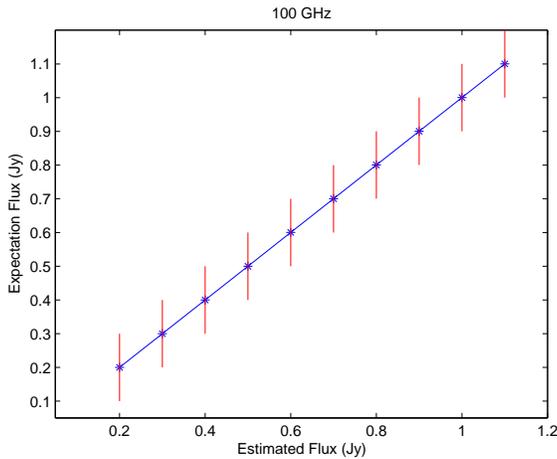}
 \caption{\label{fig9}{
 Expectation value of the flux against estimated flux. The $95\%$ confidence intervals are also plotted (100 GHz).    }}
\end{figure}

In Figure~\ref{fig9} we plot the expectation value of the flux
against the estimated flux for the $100$ GHz channel. The
expectation value is nearly the same as the most probable value. We
also plot the $95\%$ confidence intervals. In this way, we have an
idea of the uncertainty of our estimates, this confidence interval
is $\simeq 0.20$ Jy  (estimation error). The results at other
frequencies are similar.

\section{Conclusions} \label{sec:conclusions}

In this paper we propose a new strategy based on Bayesian
methodology (BM), that can be applied to the blind detection of
point sources in CMB maps. The method incorporates three prior
distributions: a uniform distribution (\ref{eq:prior_locations}) on
the source locations, an extended power law on the source fluxes
(\ref{eq:normalized_gcauchy}) and a Poisson distribution on the
number of point sources per patch (\ref{eq:prior_number}). Together
with a Gaussian likelihood, these priors produce the negative
log-posterior (\ref{eq:log_posterior}).

We minimize this negative log-posterior with respect to the source
fluxes in order to estimate them. At the same time, we show that the
detected sources must be in the peaks of the matched-filtered maps.
Finally, we choose the number of point sources which minimizes
(\ref{eq:log_posterior}) for the estimated fluxes.

In this way, we give a non-arbitrary method to select the number of
point sources. Finally, to check the performance of this technique,
we carry out flat CMB simulations for the Planck channels from 30 to
100 GHz. For simplicity, we have excluded the foregrounds in our
simulations, assuming that we are considering zones of the sky which
have been cleaned by the application of component separation
methods. However, we have included the confusion noise due to
unresolved point sources in our simulations.

We compare our Bayesian strategy with the application of a matched
filter with a standard $5\sigma$ threshold. We calculate the
contamination, the completeness and the relative error for both
methods. Though the percentage of spurious sources is a little
higher for the BM at low fluxes $ \simeq 0.2-0.3$ Jy, the
completeness is much better, allowing us to obtain catalogues with a
$99\%$ completeness and no spurious sources from $0.7$ Jy (30 GHz),
$0.8$ Jy (44 GHz), $0.55$ Jy (70 GHz) and $0.3$ Jy (100 GHz) on. The
reconstruction errors in the estimated fluxes are similarly low for
both methods.

\section*{Acknowledgments}

The authors acknowledge partial financial support from the  Spanish
Ministerio de Ciencia e Innovaci\'on project AYA2007-68058-C03-02
and from the joint CNR-CSIC research project 2008-IT-0059. KK was
supported by the Italian Space Agency, ASI, under the program on
Cosmology and Fundamental Physics, and by the TRIL program at the
Abdus Salam International Centre of Theoretical Physics, through a
specific collaboration agreement with ISTI-CNR. ES and EEK
acknowledge support from ASI through ASI/INAF Agreement I/072/09/0
for the Planck LFI Activity of Phase E2. We also thank J.
Gonz\'alez-Nuevo for useful comments and help. We wish also to thank
the referee for his comments that have helped us to improve
significantly the quality of this work.



\bibliographystyle{mn2e}
\bibliography{mibib}

\label{lastpage}

\end{document}